\documentclass[
 amsmath,amssymb,
 aps,
 prl,
 longbibliography,
 lengthcheck,%
]{revtex4-1}

\usepackage{graphicx}% Include figure files
\usepackage{dcolumn}% Align table columns on decimal point
\usepackage{bm}% bold math
\begin{document}

%\preprint{}

\title{Measurement of an Efimov trimer binding energy in a three-component mixture of $^{6}$Li}


\author{Shuta Nakajima$^{1}$}\email[Email address: ]{shuta@cat.phys.s.u-tokyo.ac.jp} \author{Munekazu Horikoshi$^{2}$, Takashi Mukaiyama$^{2,3}$, Pascal Naidon$^{2}$}
\author{Masahito Ueda$^{1,2}$}
%\thanks{}
\affiliation{$^1$\mbox{Department of Physics, University of Tokyo, 7-3-1, Hongo, Bunkyo-ku, Tokyo 113-0033, Japan}\\
$^2$\mbox{ERATO Macroscopic Quantum Control Project, JST, 2-11-16 Yayoi, Bunkyo-ku, Tokyo 113-8656, Japan}\\
$^3$Center for Frontier Science and Engineering and Institute for Laser 
Science, University of Electro-Communications, 1-5-1 Chofugaoka, Chofu, Tokyo 182-8585, Japan.}


\date{\today}

\begin{abstract}
The binding energy of an Efimov trimer state was precisely determined via radio-frequency association. 
It is found that the measurement results significantly shift with temperature, 
but that the shift can be made negligible at the lowest temperature in our experiment. 
The obtained trimer binding energy reveals a significant deviation from
the nonuniversal theory prediction based on a three-body parameter with a monotonic energy dependence.

% We performed the spectroscopy at various temperatures and confirmed that the lowest temperature condition in our experiment 
% does not suffer from any finite-temperature shift in the spectroscopy.

% The binding energy of an Efimov trimer state was measured at various temperatures via a radio-frequency association. 
% The zero-temperature extrapolation of the spectroscopic data reveals a significant deviation from the nonuniversal theory prediction 
% based on a three-body parameter with a monotonic energy dependence.
\end{abstract}

%\keywords{}

\maketitle

% \section{Introduction}

About forty years ago, V. Efimov predicted that the existence of universal trimer states known as the Efimov states, 
in a three-body system with resonant short-range interactions \cite{Efimov}. 
Such universal states are characterized only by the two-body scattering lengths for each pair of particles and a three-body parameter fixed by short-range physics.
Owing to magnetic Feshbach resonances \cite{Chin}, ultracold atomic systems turned out to be the first systems where the Efimov effect was observed conclusively.
Since the first experimental evidence in an ultracold cesium gas \cite{Kraemer},
general properties of few-body systems near unitarity such as the universal scaling laws \cite{Braaten_UT}
were confirmed in many ultracold bosonic systems \cite{Knoop,Barontini,Zaccanti,Pollack,Gross} and 
a three-component fermionic gas of ${}^6$Li \cite{Ottenstein,Huckans,Williams,Nakajima,Lompe2}, 
via the inelastic collision enhancements and minima occurring at particular intensities of an externally-applied magnetic field.
Although these features are qualitatively explained by Efimov's universal theory (UT) \cite{Braaten_UT}, 
their relative positions of loss features are shifted significantly from universal predictions. 
For example, the shift of the atom-dimer loss peaks from that expected from the three-body loss peaks \cite{Knoop,Nakajima,Lompe2}
%about 50\% 
and the notable discrepancies in properties of the Efimov resonances between regions of positive and negative scattering lengths \cite{Pollack} 
do not seem to be consistent with a fixed three-body parameter.
Therefore, the precise determination of the three-body parameter is crucial to understand these systems. 
% Efforts are currently being made to understand these shifts.
% Especially, we revealed that shift of the atom-dimer resonance could not be explained by two-body scattering length with finite range corrections 
% and constant three-body parameter, and construct non-universal models with "energy-dependent" three-body parameter \cite{Nakajima}. 
% Although we could indirectly check the validity of our model from the atom-dimer loss measurement in the different combination of atoms and dimers \cite{Lompe2, Naidon_rev}, 
% the measurement of the binding energy of the Efimov states have been wanted to check the validity of the model directly.

% In the previous paper, we first constructed nonuniversal model by taking into account the energy dependence of the scattering length which accounts for the two-body finite-range corrections to explain the atom-dimer loss feature in the three-component gas of $^6$Li.
To understand the atom-dimer loss feature in the three-component gas of $^6$Li, we constructed a nonuniversal model by taking into account the energy dependence of the
scattering length due to finite-range corrections \cite{Nakajima}.
This two-body physics correction still does not explain the atom-dimer loss feature that we observed experimentally.
We then introduced an energy-dependent three-body parameter $\Lambda$ which phenomenologically reproduces all the experimental data of the three-body loss and the atom-dimer loss in the three-component mixture of $^6$Li atoms.
\cite{Nakajima,Naidon_rev}.
However, these three-body and atom-dimer loss measurements provide information on $\Lambda$ only at the points where
the trimer energy level vanishes upon dissociation or meets a dimer energy level.
% (namely, the locations of the peaks in the loss rate coefficient). 
Thus, it has been desirable to directly measure the binding energy of the Efimov trimers to fully determine the three-body parameter and the validity of the model.

Recently, T. Lompe {\it et.al.}\cite{Lompe} demonstrated a radio-frequency (RF) association of the Efimov trimer state in the three-component mixture of $^6$Li atoms.
This method constitutes the most direct observation of Efimov trimers so far, and provides a way to determine the binding energy of the Efimov trimers 
not only at particular points, but also over a continuous range of the magnetic field. 
The measured energies reported in \cite{Lompe} are seemingly in very good agreement with the predictions of our phenomenological model \cite{Nakajima}. 
However, their measurement was done at a temperature of 1 $\mu$K which can shift the resonance of the RF spectroscopy on the order of 30~kHz \cite{Lompe}. 
To precisely determine the three-body parameter from the binding energy measurement, it is important to take into account the temperature effect in the RF spectroscopy.
 
%%%%%%%%%%%%%%%%%%%%%%%%%%%%%%%%%%%%%%%%%%%%%%%%%%%%%%%%%%%%%%%%%%%%%%%%%%%%%%%%%%%%%%%%%%%%%%%%%%%%%%

In this Letter, we report on the measurements of the binding energy of the Efimov trimer state and the precise determination of the three-body parameter.
% in three-component mixture of $^6$Li via RF association
In particular, we observed the temperature dependence of the RF spectrum, and found a resonance shift with temperature. 
Operating at lower temperature to eliminate this shift, we found that the measured energies significantly deviate from those of Ref. \cite{Lompe} 
and the previous predictions \cite{Nakajima}. 
Refining the model to fit these measurements, we obtain a non-monotonic energy dependence of the three-body parameter $\Lambda$.

% precisely measured  in three-component mixture of $^6$Li  with very low collisional energy.
% We checked the temperature dependence of the RF spectrum and took all spectrum at 70 nK to remove the effect of the temperature shift.
% We found that the measured values at low temperature were significantly deviated from previous experiment \cite{Lompe} 
% and our own predictions based on previous experiments\cite{Nakajima}, as well as universal theory (UT) predictions \cite{Braaten_Li}.
% We derived the three-body parameter $\Lambda $ from the measured binding energies and discussed the new energy-dependent behaviour.

%%%%%%%%%%%%%%%%%%%%%%%%%%%%%%%%%%%%%%%%%%%%%%%%%%%%%%%%%%%%%%%%%%%%%%%%%%%%%%%%%%%%%%%%%%%%%%%%%%%%%%%%

% We performed our experiments as follows.
We use a mixture of fermionic $^6$Li atoms in the lowest three hyperfine states 
$|F;m_F\rangle= |1/2;1/2\rangle$, $|1/2;-1/2\rangle$ and $|3/2;-3/2\rangle$, which we label as $|1\rangle$, $|2\rangle$ and $|3\rangle$, respectively. 
Because of their fermionic nature, only distinguishable particles interact via s-wave scattering, i.e., pair interactions are described by three different two-body scattering lengths $a_{12}$, $a_{23}$ and $a_{31}$, which diverge at 834~G, 811~G, and 690~G respectively due to Feshbach resonances.
These three broad and overlapping Feshbach resonances allow us to precisely tune all three scattering lengths simultaneously via magnetic field.
At first, we create a degenerate Fermi gas of $^6$Li atoms in the two lowest hyperfine states $|1\rangle$ and $|2\rangle$ as described in detail in \cite{Inada}. 
Then we evaporatively cooled the atoms with a population ratio of $|1\rangle:|2\rangle=1:2\sim1:4$ at 834~G, 
and adiabatically ramped down the magnetic field to 685~G-740~G in 300~ms. At this field ramp, $|12\rangle$ dimers, which were formed from associating states $|1\rangle$ and $|2\rangle$, 
were adiabatically formed and excess atoms in $|2\rangle$ still remained in the trap. 
In this way, we obtained a mixture of $|2\rangle$ atoms and $|12\rangle$ dimers.

% The temperature was controlled by the final trap depth of the forced evaporative cooling process.
In our experiment, temperature was controlled by changing the depth of the optical trap. For the measurement at high temperatures ($T > 500$ nK), 
we used a single-beam optical trap with a beam waist of 33~$\mu$m (we call this a ``tight trap"). 
For measurements at lower temperatures ($T < 500$ nK), we used a large-volume hybrid magnetic/optical trap with an effective beam waist of 300~$\mu$m (``shallow trap"). The trap frequencies for the tight (shallow) trap were approximately given by
% $\omega_x/2\pi = 67(1)\sqrt{P}$ Hz (54(1) Hz),
% $\omega_y/2\pi = 52(2)\sqrt{P}$ Hz (44(1) Hz) and 
% $\omega_{z}/2\pi = \sqrt{0.26(4)B+0.20P}$ Hz ($\sqrt{0.309(2)B}$ Hz) 
$\omega_x/2\pi = 67\sqrt{P}$ Hz (54 Hz),
$\omega_y/2\pi = 52\sqrt{P}$ Hz (44 Hz) and 
$\omega_{z}/2\pi = \sqrt{0.26B+0.20P}$ Hz ($\sqrt{0.31B}$ Hz) 
in the $x$, $y$ and $z$ directions, respectively, where $P$ is the power of the optical dipole trap in mW and $B$ is the strength of the magnetic field in Gauss.
The laser power for the tight trap was varied between 80 and 660~mW, which corresponds to a temperature from 500~nK to 2~$\mu$K.
The total number of atoms in state $|2\rangle$ before creating dimers was $10^6 - 10^5$ depending on the final temperature.

%%%%%%%%%%%%%%%%%%%%%%%%%%%%%%%%%%%%%%%%%%%%%%%%%%%%%%%%%%%%
\begin{figure}[tbp]
\includegraphics[scale=0.45]{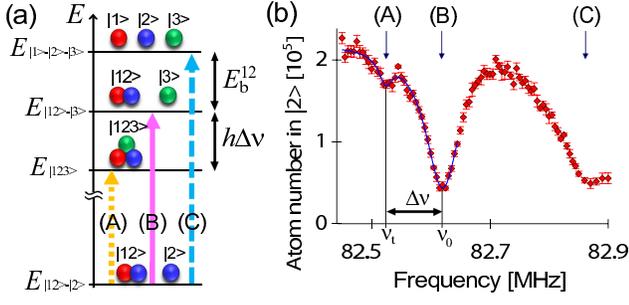}
\caption{\label{fig:Eb} (Color online)  
(a)Energy levels and transitions associated with the RF association of trimers. 
Transitions (A), (B) and (C) correspond to an association of Efimov trimer, bare atomic transition between $|2\rangle$ and $|3\rangle$
and dissociation of $|12\rangle$ dimers, respectively.
(b)RF spectrum at 700~G. Dips (A), (B) and (C) correspond to signals of the transition (A), (B) and (C), respectively.
Each data point is the average of 3 - 5 measurements.
}
\end{figure}%
%%%%%%%%%%%%%%%%%%%%%%%%%%%%%%%%%%%%%%%%%%%%%%%%%%%%%%%%%%%%

Figure \ref{fig:Eb} (a) represents the energy levels and transitions related to the RF association of trimers.
We started from a mixture of $|2\rangle$ atoms and $|12\rangle$ dimers at the fields of interest and applied an RF field at frequencies around the transition from $|2\rangle$ to $|3\rangle$.
We applied the RF fields to the atom-dimer mixture for 30~ms (highest temperature) $-$ 300~ms (lowest temperature) using an antenna designed to be resonant at 80~MHz. 
% driven by a 100W RF-amplifier. 
The Rabi frequency was 7 kHz for the $|2\rangle$-$|3\rangle$ transition at 705~G.
After RF-pulse we performed state-selective absorption imaging at 834~G, and counted the number of atoms in $|2\rangle$ after dissociating $|12\rangle$ dimers.

Figure \ref{fig:Eb} (b) shows a typical RF spectrum taken at 700~G at the temperature of 70 nK. %70*(1.4/1.2)^2
The central dip (B) at $\nu_0$ corresponds to the bare atomic transition between $|2\rangle$ and $|3\rangle$. The width of the transition reflects the inelastic collision rate between $|3\rangle$ atoms and $|12\rangle$ dimers \cite{power_broadening}. 
The left dip (A) at $\nu_t$ corresponds to the association to Efimov trimers. 
Association occurs when the detuning from the bare atomic transition $\Delta \nu=\nu_0-\nu_t$ matches to the difference between 
the $|12\rangle$ dimer binding energy $E_b^{12}$ and the trimer binding energy $E_{|123\rangle}$, i.e. $\nu_t=\nu_0 -E_{|123\rangle}/h+E_b^{12}/h$,
where $h$ is Planck's constant.
The right dip (C) is the signal from the dissociation of $|12\rangle$ dimers.
The location of the bare atomic transition and the dip of the Efimov association were determined by fitting the data with a double-Lorentzian function, $N(\nu)=N_0-A_0/[1+(\nu-\nu_0)^2/(\Gamma_0/2)^2]-A_t/[1+(\nu-\nu_t)^2/(\Gamma_t/2)^2]$.
The free parameters are amplitude, position and width of the free-free transition ($A_0, \nu_0, \Gamma_0$) and those of trimer association dips ($A_t, \nu_t, \Gamma_t$), and offset $N_0$. 
% We confirmed that a double-Gaussian function fitting also gives the same dip position within the fitting errors.

%%%%%%%%%%%%%%%%%%%%%%%%%%%% Temperature dependence %%%%%%%%%%%%%%%%%%%%%%%%%%%%%%%%%%%%%%%%%%%%%%%%%%%

%%%%%%%%%%%%%%%%%%%%%%%%%%%%%%%%%%%%%%%%%%%%%%%%%%%%%%%%
\begin{figure}[tbp]
\includegraphics[scale=0.45]{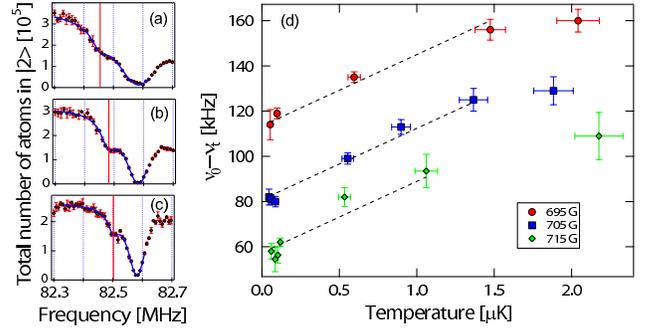}
\caption{\label{fig:temp_dep} (Color online) (a)-(c) Temperature dependence of the RF association spectrum at 705 G. 
The data were taken at 1370(90)~nK (a), 554(37)~nK (b) and 86.5(4.2)~nK (c).
The Efimov association peak position (vertical red line) shifts as temperature changes. 
(d) Temperature dependence of the peak difference at 695 G (red circles), 705 G (blue squares) and 715 G (green diamonds).
The black dashed lines show $1.5 k_{\rm B}T$ dependence.
}
\end{figure}%
%%%%%%%%%%%%%%%%%%%%%%%%%%%%%%%%%%%%%%%%%%%%%%%%%%%%%%%%

When the kinetic energies of the atoms and dimers are on the same order as the resolution of the RF spectroscopy, 
we need to take into account the resonance shift due to finite temperature effects. 
Figures \ref{fig:temp_dep} (a)-(c) show the temperature dependent RF spectra taken at 705 G. 
We can clearly see that the dip of the Efimov resonance shifts to the left as the temperature increases. 
Here we estimate the temperature of the cloud from the time-of-flight images of $|2\rangle$ atoms at zero magnetic field. 
The magnetic field was turned off simultaneously with the optical trap to suddenly eliminate the interaction between atoms and dimers, 
and the expansion of $|2\rangle$ atoms reflects the temperature of the cloud. 
Although $T/T_{\rm F}$ is around 0.5 at the lowest ($T_{\rm F}$ is the Fermi temperature of the trapped system), 
the collisional energy distribution remains close to a Boltzmann distribution \cite{BD}.
% Although $T/T_{\rm F}$ is around 0.5 at lowest ($T_{\rm F}$ is the Fermi temperature of the trapped system), 
% the collisional energy distribution did not deviate from the Boltzmann distribution so much \cite{BD}.
% the momentum distribution is almost the same as a Boltzmann distribution \cite{BD}.

Figure \ref{fig:temp_dep} (d) shows the temperature dependence of the Efimov resonance location measured from the bare atomic transition $\nu_0 - \nu_t$ at 695~G, 705~G and 715~G. We found that the shift due to temperature below 1.5 $\mu$K is well described by $1.5 k_{\rm B} T$ (dashed lines)
as expected from the Boltzmann distribution \cite{BD2}. 
It is also expected that the slope of the temperature dependence becomes smaller when $k_{\rm B}T/h$ is comparable with the transition linewidth. 
A change in the slope can be seen at around 1.5~$\mu$K in Fig. \ref{fig:temp_dep} (d), which seems to suggest that the linewidth of the Efimov trimer state in this magnetic field region is roughly 1.5~$\mu$K $\sim$ 30~kHz, which is consistent with the fitting result of $\Gamma_t$. 
The density of the $|2\rangle$ atoms is $1 \times 10^{11}$ cm$^{-3}$ for the shallow trap (data point for $T<0.5$~$\mu$K) and $2 \times 10^{12}$ cm$^{-3}$ for the tight trap. 
We estimate that the resonance shifts due to collisions between trimers and atoms or dimers are negligible in the shallow trap corresponding to the low temperature measurements.
In the tight trap, we noticed a shift of the bare atomic transition resonance on the order of 10(2)~kHz, which is within the accuracy of our measurement.
% Although it is not clear whether interaction between trimers and atoms or dimers affect resonance position in the RF spectra, 
% the linear dependence of the resonance shift as a function of temperature shown in Fig. \ref{fig:temp_dep} (d) 
% seems to indicate that the interaction effect does not play an important role in the present measurement. 

% In the shallow trap the collisional shifts are negligible because the density is low.
% In the end, we only use the results in the shallow trap for the trimer energy, so I think
% the reader just wants to know if the final results can be affected by collisional shifts or not.
Note that the measurement in \cite{Lompe} and our data points around 1~$\mu$K in Fig. \ref{fig:temp_dep} (d) are consistent. 
% In their paper, they compare their result with our calculation based on the energy dependent three-body parameter and showed good agreement. 
However, since there is a non-negligible shift due to the temperature effect as shown in the present measurement, 
we need to extrapolate the plot in Fig. \ref{fig:temp_dep} (d) toward zero temperature to extract the actual binding energy from the RF spectra. 
% The lowest temperature we obtained in our experiment is 60 - 70 nK (left most point in Fig. \ref{fig:temp_dep} (d)), 
% which almost gives the same result with zero temperature. Therefore we use this experimental condition to determine the binding energy of trimers.

% The difference between the upper bound (lower bound) of the predictions and our lowest temperature data at 715~G is 46(3)~kHz (30(3)~kHz), 
% which corresponds a 17(1) \% (12(1) \%) deviation from the binding energy of the trimer calculated from previous nonuniversal model.
% And note that the difference between predictions and our lowest temperature data at 695~G is 29(7)~kHz (14(7)~kHz), 
% which corresponds 7.0(1.6) \% (3.6(1.7)\%) deviation from upper bound (lower bound) of the predictions.

%%%%%%%%%%%%%%%%%%%%%%%%%%%% Low Temperature measurement %%%%%%%%%%%%%%%%%%%%%%%%%%%%%%%%%%%%%%%%%%%%%%%%%%%55

\begin{figure}[tbp]
\includegraphics[scale=0.48]{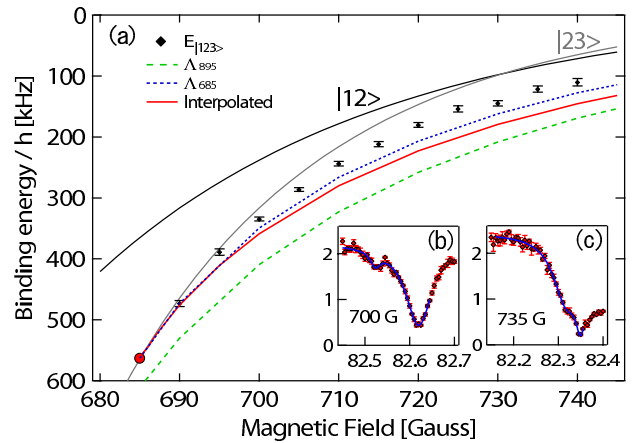}
\caption{\label{fig:Efimov_Eb} (Color online) (a) Magnetic-field dependence of the binding energy of the Efimov trimer states.
The blue dotted (the green dashed) curve indicates calculated binding energy using constant 3-body parameter $\Lambda_{685}$ ($\Lambda_{895}$).
The red solid curve is given by our previous nonuniversal theory with energy dependent 3-body parameter 
assuming monotonic energy dependence and adjusted to match both resonances at 685~G and 895~G \cite{Nakajima}.
Measured binding energies at low temperature is indicated with black diamond and its  error bars include fitting uncertainties, 
uncertainty of the calculated dimer binding energy (2~kHz) and temperature shift estimated from $1.5 k_{\rm B}T$.
The red point indicates the position of the observed $|1\rangle-|23\rangle$ atom-dimer loss peak \cite{Nakajima, Lompe2}. %and three-body loss peak \cite{Huckans}.
The solid curves labeled $|12\rangle$ and $|23\rangle$ are the binding energies of dimer $|12\rangle$ and $|23\rangle$, respectively.  
Insets (b) and (c) show the RF spectra at 700~G and 735~G respectively. 
The units of the inset figures are the same as that of Fig.\ref{fig:Eb} (b). 
}
\end{figure}

Figure \ref{fig:Efimov_Eb} (a) shows the magnetic-field dependence of the binding energy divided by $h$. All data points were taken at 70 nK. The binding energy of the trimer is given by $E_{|123\rangle}=E_b^{12}+(h \nu_0-h \nu_t)$, where $E_b^{12}$ is the binding energy of the $|12\rangle$ dimer from coupled-channel equations using singlet and triplet potentials for $^6$Li \cite{Julienne}.  
The separation between the Efimov association dip and atom resonance become smaller as the magnetic field increases. 
To resolve these two resonances, we made the bare atomic transition narrower by reducing the number of dimers.
Since the linewidth of the bare atomic transition is determined by the inelastic collision rate of $|3\rangle$ atoms with $|12\rangle$ dimers, reducing the number of $|12\rangle$ dimers makes the atomic resonance narrower and helps separate the Efimov resonance from the atomic resonance. 
Above 715~G where the Efimov resonance frequency gets closer to the atomic resonance, we used imbalanced atom-dimer mixtures whose population ratio is about $|12\rangle:|2\rangle\sim 1:3$.
We confirmed that changing the population ratio does not change the position of the association dips. In this way, we were able to determine the binding energy of Efimov trimers up to 740~G \cite{fit_error}. We could not observe an association dip at and below 685~G. This is consistent with the previous atom-dimer loss experiments \cite{Nakajima,Lompe2} in which the dimer-trimer meeting point was observed at 685~G.

% pointed out
% because of the sensitibity... the variation of the 

%%%%%%%%%%%%%%%%%%%%%%%%%%%%%%%%%%%%%%%%%%%%%%%%%%%%%%%%%%%%
%%%%%%%%%%%%%%%%%%%%%%%%%%%%%%%%%%%%%%%%%%%%%%%%%%%%%%%%%%%%

In previous works \cite{Nakajima,Naidon_rev}, we showed that the universal model cannot accurately describe the trimer because the dimer binding energy is already off by 8 \% from its universal behaviour. 
Taking into account the non-universal two-body behaviour using energy-dependent scattering lengths, 
we determined the two values of the three-body parameter $\Lambda_{\rm685}$ and $\Lambda_{895}$ 
which reproduce the observed dimer-trimer meeting point at 685 G \cite{Nakajima, Lompe2}, and the trimer dissociation point at 895 G \cite{Williams}, respectively.  
These two values are not consistent and lead to two different trimer energy curves indicated by the blue dotted and green dashed curves in Fig. \ref{fig:Efimov_Eb}. 
This lead us to the conclusion that the three-body parameter $\Lambda$ must vary, presumably with energy. 
We made an interpolation of $\Lambda$, leading to the trimer energy curve indicated by the red solid curve in Fig. \ref{fig:Efimov_Eb}. 
Fortuitously, the trimer energy reported in \cite{Lompe} from 1~$\mu$K measurements agree with this interpolated curve. 
However, our new data do not follow this curve, and suggest a more complicated variation of $\Lambda$.
% the deviation of the measured binding energy from our nonuniversal theory prediction is clearly seen since we now have more experimental data points to compare with theory. Considering that 685~G is one of the points of the boundary condition, it is natural that the deviation between data and theoretical prediction becomes smaller when the magnetic field is closer to 685~G. 

%%%%%%%%%%%%%%%%%%%%%%%%%%%%%%%%%% Theory part %%%%%%%%%%%%%%%%%%%%%%%%%%%%%%%%%%%%%

%%%%%%%%%%%%%%%%%%%%%%%%%%%%%%%%%%%%%%%%%%%%%%%%%%%%%%%%%%%%
\begin{figure}[tbp]
\includegraphics[scale=0.54]{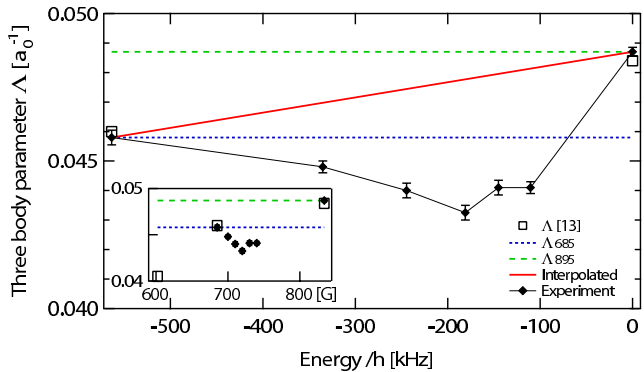}
\caption{\label{fig:theory} (Color online) Energy-dependence of the three-body parameter $\Lambda$. 
The black open squares show $\Lambda$ previously determined from three-body loss \cite{Williams} and atom-dimer loss \cite{Nakajima}.
The blue dotted line and the green dashed line show constant three-body parameters $\Lambda_{685}$ and $\Lambda_{895}$, respectively.
The red curve indicates $\Lambda$ for our interpolated curve in Fig. \ref{fig:Efimov_Eb}.
The black diamonds show the modified energy-dependence of $\Lambda$ to reproduce the measured binding energy.
Insets shows the magnetic-field dependence of the three-body parameters. 
}
% The error bars for these points arise from experimental uncertainties.}
\end{figure}
%%%%%%%%%%%%%%%%%%%%%%%%%%%%%%%%%%%%%%%%%%%%%%%%%%%%%%%%%%%%

It should be noted that the value of $\Lambda$ is very sensitive to the two-body model, in particular the uncertainty of the scattering length. 
For example, the experimental uncertainties of dimer binding energy of $\sim$2~kHz at 720~G reported in \cite{Bartenstein} 
gives an uncertainty of at most $\sim$ 0.7 \% in the scattering lengths
but it can cause a 5 \% variation of $\Lambda$. As such, the value of $\Lambda$ per se is not so meaningful. 
As long as the deviation of the scattering length in our model from the real one is a smooth nearly-constant shift, 
the energy dependence of the three-body parameter shows the same behavior.
Therefore we can investigate the variation of $\Lambda$.
% However, assuming that the variation of the scattering length in our model is essentially the same as the real one, 
% up to some smooth nearly-constant shift over the experimental range of magnetic field, we can investigate the variation of $\Lambda$. 
For a given two-body model \cite{Julienne}, we adjust the value of $\Lambda$ to reproduce the trimer energy for each experimental point. 
The resulting variation of $\Lambda$ is indicated by black diamonds in Fig. \ref{fig:theory}. 
Curiously enough, the variation of $\Lambda$ does not seem monotonic. 
This suggests that the naive expectation that the three-body parameter should have a simple nearly constant behaviour may not be true in such atomic Feshbach
resonant systems.
Understanding such a peculiar variation using models beyond the zero-range single-channel approximation constitutes a new challenge. 
For example the three-body interaction between the small closed-channel molecular component and the third atom \cite{Yamashita} may contribute to such a variation of the three-body parameter. 
Our preliminary calculations show that very large interactions could shift the energy by as much as 20~kHz.

%%%%%%%%%%%%%%%%%%%%%%%%%%%%%%%%%%%%%%%%%%%%% discussion %%%%%%%%%%%%%%%%%%%%%%%%%%%%%%%%%%%%%%%%%%%%%

In summary, we have measured the binding energy of an Efimov state by RF association in a three-component mixture of $^{6}$Li atoms. 
% over a range of magnetic field of 50 G. 
We found that the observed RF association dip shifts with temperature and performed our association at very low temperature ($\leq 70$ nK). 
Our lowest temperature measurements reveal that the binding energy of the Efimov states is smaller than expected. 
Our results suggest a peculiar variation of the effective three-body parameter which sets the trimer binding energy in zero-range models. 
This new information on the non-universal Efimov physics of three-component $^6$Li constitutes a new challenge for few-body theories.
% Although we could not fully understand this nonuniversal behavior, our results provided new information on the Efimov physics of three-component $^6$Li
% which constitutes a new challenge for few-body theories.
% for understanding of the Efimov physics in three-component $^{6}$Li and will be a strong "touchstone" of few-body theories. 
A full understanding of the Efimov spectrum will provide quantitative explanations for other experiments on the three component mixtures of $^6$Li 
such as the observations of loss minima in atom-dimer mixtures of $|12\rangle$ and $|3\rangle$ \cite{Lompe2}, 
which have received so far only qualitative \cite{Hammer2} or phenomenological \cite{Naidon_rev} explanations.
% which have not yet been explained quantitatively by neither universal theory \cite{Hammer2} nor our previous theory \cite{Naidon_rev}.     
 
\begin{acknowledgments}
We thank A. Wenz, T. Lompe, Y. Castin and L. Pricoupenko, P. S. Julienne, E. Tiesinga and A. Simoni for useful discussions. 
This work was supported by KAKENHI (22340114, 22103005), 
Global COE Program "the Physical Sciences Frontier" and the Photon Frontier Network Program, MEXT, Japan.
S.N. acknowledges support from the Japan Society for the Promotion of Science.
\end{acknowledgments}

% \bibliography{basename of .bib file}

\begin{thebibliography}{99}
\bibitem{power_broadening} We confirmed that the width of free-free transiton decreases as the number of dimers decreases with the same driving RF power.
\bibitem{BD} The deviation of kinetic energy of a classical gas and an ideal Fermi gsa in a harmonic trap is 6\% at 0.5~$T_F$.
Therefore our TOF temperature could shift about 6~\% from the real temperature, which is within our temperature uncertainty.
\bibitem{fit_error} We could have additional systematic errors due to our simplistic model for the lineshapes. 
The double-Lorentzian fitting might have problems in the region where the overlap between the free-free resonance and Efimov association resonance is large, 
i.e., there may be systematic errors as large as 10 kHz.
\end{thebibliography}

%\textbf{Acknowledgements}\\
%The authors acknowledge ... for comments and discussions.

\end{document}